\documentclass[prb,nopacs,nobibnotes,showkeys,superscriptaddress]{revtex4}
\usepackage{amsmath,amssymb,epic,eepic}
\usepackage{isolatin1}
\usepackage{graphicx}
\usepackage{hyperref}

\begin{document}


\title{Electron quantum optics in quantum Hall edge channels}

\author{Charles Grenier}

\address{Laboratoire de Physique, Ecole Normale Supérieure de Lyon\\
46 allée d'Italie, 69007 Lyon, France}

\author{Rémy Hervé}

\affiliation{Laboratoire de Physique, Ecole Normale Supérieure de Lyon\\
46 allée d'Italie, 69007 Lyon, France}

\author{Gwendal Fève}

\affiliation{Laboratoire Pierre Aigrain, Ecole Normale Supérieure\\
24 rue Lhomond, 75005 Paris, France}

\author{Pascal Degiovanni}\thanks{To appear in the proceedings of StatPhys 24 satellite conference on {\em International Conference on Frustrated Spin Systems,
Cold Atoms and Nanomaterials} held in Hanoi (14-16 July 2010).}

\affiliation{Laboratoire de Physique, Ecole Normale Supérieure de Lyon\\
46 allée d'Italie, 69007 Lyon, France}

\begin{abstract}
In this paper, we review recent developments in the emerging field of electron quantum optics, stressing analogies and
differences with the usual case of photon quantum optics. Electron quantum optics aims at preparing, manipulating and measuring coherent
single electron excitations propagating in ballistic conductors such as the edge channels of a 2DEG in the integer quantum Hall regime. 
Because of the Fermi statistics and the presence of strong interactions, electron quantum optics exhibits new features compared to 
the usual case of photon quantum optics. In particular, it provides a natural playground to understand decoherence and 
relaxation effects in quantum transport.
 \end{abstract}


\maketitle

\section*{Introduction}

The development of quantum optics started in the 50s and 60s, motivated by the need to understand photodetection, 
statistics and coherence of light within the framework of quantum theory. 
In the late 50s, the seminal Hanbury-Brown and Twiss (HBT) experiment\cite{Hanbury:1956-1,Hanbury:1956-2} demonstrated 
the bunching of photons emitted by 
a classical chaotic source and stimulated the development of the quantum theory of optical coherences 
by Glauber\cite{Glauber:1962-1,Glauber:1963-1,Glauber:1963-2}. 
An important milestone in quantum optics was the demonstration of the non-classical character of 
fluorescence radiation emitted by a single atom\cite{Kimble:1977-1}
and, later, the demonstration of photons sources showing photon antibunching\cite{Hong:1986-1,Grangier:1986-1}. 
As can be seen from a recent review\cite{Lounis:2005-1}, a variety of single photon sources
have been demonstrated: some of them are based on single molecules \cite{Brunel:1999-1} or colored centers in diamond\cite{Beveratos:2002-1} and others
on artificial nanostructures such as quantum wells\cite{Kim:1999-1} or quantum dots in microcavities\cite{Pelton:2002-1}. 
Such sources are used to implement quantum key distribution protocols. An important effort is now oriented towards reliable high rate 
sources of entangled photon pairs\cite{Yamamoto:2003-1,Kuzmich:2003-1} in the perspective of quantum communication protocols\cite{Kok:2007-1}.
Since Glauber's seminal papers, a lot of work has been devoted to the production and characterization
of quantum states of the electromagnetic field. In particular, spectacular results in the preparation and measurement of the quantum
state of an electromagnetic field in the microwave domain have been achieved in cavity QED, either using Rydberg atoms in ultrahigh finesse superconducting
cavities\cite{Book:Haroche-Raimond} or using superconducting nanocircuits. In this latter stream of research, known as circuit QED\cite{Blais:2004-1}, recent progresses
have been spectacular, including for example the generation of arbitrary quantum states up to 12 photons in a cavity\cite{Hofheinz:2009-1}.

\medskip

Following this historical trend, progresses in mesoscopic physics have increasingly brought 
closer the dream of performing quantum optics like experiments with electrons in condensed 
matter systems where many body effects are expected to play a crucial role. During the last twenty 
years, electronic interferences observed in conductance measurements routinely provided information 
about first order electron coherence, although the use of quantum optics language was not emphasized.  This 
recently culminated with Mach-Zehnder experiments performed on edge channels 
of high mobility 2DEG in the integer quantum Hall effect regime\cite{Ji:2003-1}, showing large coherence lengths 
in excess of 10~$\mu$m at low temperature\cite{Roulleau:2008-2}. 

Following a parallel path with optics, in the beginning of the 90s, theoreticians proposed to investigate intensity 
fluctuations of the electrical current\cite{Lesovik:1989-1,Buttiker:1990-1,Martin:1992-1,Buttiker:1992-1} such as shot noise, 
a manifestation of electron granularity first pointed out by W. Schottky\cite{Schottky:1918-1}. The analogy between electronic 
quantum shot noise and photon noise was soon established and the importance of studying current correlations
in the spirit of Hanbury Brown and Twiss became evident\cite{Martin:1992-1,Buttiker:1992-1}. In 2007, following a 
proposal by Samuelsson, Sukhorukov and B\"uttiker\cite{Samuelsson:2004-1}, Heiblum and 
his collaborators demonstrated two particle interferences in quantum Hall edge channels\cite{Neder:2007-2}

\medskip

These considerations emphasized the interest for realizing ``electron quantum optics'' experiments 
but it is only recently that the fine control of electron dynamics and high frequency current noise measurements at relevant 
time-scales have been achieved\cite{Zakka-bajjani:2007-1}. A further breakthrough has been 
brought recently with the experimental realization of an on demand single electron source\cite{Feve:2007-1,Mahe:2010-1}. By opening the way to
experiments involving single electrons instead of the flow continuously emitted from a voltage biased contact, this source 
offers totally new opportunities. 

In this context, an important and natural question is to understand how far the analogy between ``electron quantum optics'' and 
photonic quantum optics can go? The Fermi statistics of electrons is expected to bring in new features. First of all, the 
ground state of a metallic conductor is a Fermi sea which, as we shall see, radically differs from the electromagnetic vacuum. 
Moreover, even in the absence of interactions, Fermi statistics leads to radically different phenomenon than in optics: for example, with fermions,
entanglement production from sources at equilibrium is possible even in the absence of
interactions whereas this is prohibited for bosonic particles such as photons\cite{Beenakker:2006-1}.  

Besides quantum statistics, Coulomb interactions have strong consequences on the propagation and coherence 
of electronic excitations within a ballistic conductor. Although one expects the Pauli exclusion principle to block inelastic processes at low
energies, the quantum optics paradigm of freely propagating particles is seriously questioned in electronic systems.
A crucial question is thus not only to understand the effects of quantum statistics, but also 
deviation from the non-interacting picture due to electron-electron interactions and to decoherence 
induced by the electromagnetic environment\cite{Devoret:1990-1,Girvin:1990-1,Ingold:1992-1}. 

\medskip

Here we review some of the recent developments in the emerging field of electron quantum optics, stressing analogies and
differences with the case of quantum optics with photons. Although the ultimate goal of electron quantum optics would be
to study and manipulate in a controlled way the quantum state of electrons in conductors, it should be remembered that, compared to the
usual case of photon quantum optics, this field is still very young. Although several groups are now working in this field,
only a few electron quantum optics experiments have been performed compared to what has been achieved in photon quantum optics. 
Thus, even if this review mainly focuses on theoretical aspects, we have chosen to restrict ourselves to the 
discussion of topics directly related to forthcoming experimental projects which
include the analogous of the famous Hanbury Brown \& Twiss and Hong, Ou, Mandel experiments as well as quantitative studies of electron decoherence in
quantum Hall edge channels.

\medskip

In section \ref{sec:SPC}, the basic notions needed to characterize electron quantum coherences are introduced and
the parallel with quantum optics is made explicit. Then, in section \ref{sec:HBT}, we discuss how single electron coherence can be measured thus
leading to a quantum tomography protocol for single electrons in quantum Hall edge channels. Finally, in section \ref{sec:decoherence}, we discuss 
electron decoherence and relaxation making an explicit and somehow surprising connexion between this problem and high frequency electric transport.

\section{Electron coherence}
\label{sec:SPC}

So far, experiments have been performed using phase-coherent ballistic conductors realized using high mobility 2D 
electron gases in the integer quantum Hall regime. Then, electron propagation takes place within
chiral edge channels\cite{Halperin:1982-1} suitable to mimic quantum optics setups\cite{Buttiker:1988-1}.
 Having in mind these experimental realizations, our discussion will be focused on chiral edge channels. 
Although this may sound restrictive, we stress that these systems
already provide many of the components needed for quantum optic experiments: sample edges are perfect mirrors whereas
the quantum point contact\cite{vanWees:1988-1} provides the analogous of a beam splitter for electrons\cite{vanWees:1991-1}. 
Moreover, the single electron source has been recently demonstrated\cite{Feve:2007-1,Mahe:2010-1}
and powerful noise measurement techniques have been developed\cite{Glattli:2009-1,Parmentier:2010-1}.

\subsection{From photons to electrons}

In quantum optics, quantum coherence properties of photons are described by Glauber correlation functions which are closely related to
photodetection signals\cite{Glauber:1962-1,Glauber:1963-1}. 
Photodetection is the conversion of a photon into an electrical signal. An old example is provided by the photomultiplier
which relies on the photoelectric effect to convert a single photon into a single electron and then amplifies the corresponding current
using secondary emission of electrons in dynodes.  Modern photodetectors in the optical domain are photodiodes among which single
photon avalanche photodiodes. These solid state photodetectors reach single photon sensitivity and have a photon 
arrival time resolution of a few tens of picoseconds\cite{Cova:1996-1}.

Using time dependent perturbation theory, the photodetection signal of a single photodetector at position $x$ can be expressed 
in term of two correlation functions:
\begin{equation}
I_{D}(t)=\int \mathcal{G}_{\rho}^{(1)}(x,\tau;x,\tau')\,K_{D}(\tau,\tau')\,d\tau,d\tau'\,
\end{equation}
where the first order correlation function $\mathcal{G}_{\rho}^{(1)}(x,\tau;x,\tau')$ is relative to the electromagnetic
field and $K_{D}(\tau,\tau')$ relative to the detector and encoding its properties (bandwidth, efficiency etc). 
The first order correlation function defined by Glauber\cite{Glauber:1963-1} is defined as
\begin{equation}
\mathcal{G}_{\rho}^{(1)}(x,\tau;x',\tau')=\mathrm{Tr}(E^{(+)}(x,\tau)\,\rho\,E^{(-)}(x',\tau'))
\end{equation}
where $E^{(+)}$ (resp. $E^{(-)}$) respectively denote the annihilation and creation part of the electromagnetic field and
$\rho$ its density operator. Correlators involving a single time
measure spatial coherence properties whereas taking all operators at the same position but various times gives
a measurement of temporal coherence.

A broadband detector is characterized by $K_{D}(\tau,\tau')\sim\delta(\tau-\tau')$. In this case, the photodetection signal is directly 
proportional to the instantaneous photocount $I(x,t)\sim\langle E^{(+)}(x,t)E^{(-)}(x,t)\rangle_{\rho}$ at the detector's position $x$. 
On the contrary, a narrow band detector
will only retain the contribution of photons at a given frequency: $K_{D}(\tau,\tau')\sim e^{i\Omega(\tau-\tau')}$. 
Finally, Glauber defines higher order correlators which give the photodetection signals associated with the detection coincidences by
several detectors\cite{Glauber:1963-1}. 

\subsection{Single electron coherence}

The analogous of Glauber correlators for electrons propagating in a metallic system
are electron and hole Keldysh correlation functions:
\begin{subequations}
\begin{eqnarray}
\label{eq:SPC:1-e}
\mathcal{G}^{(e)}_{\rho}(x,t;y,t') & = & \mathrm{Tr}(\psi(x,t)\rho\,\psi^\dagger(y,t'))\\
\mathcal{G}^{(h)}_{\rho}(x,t;y,t') & = & \mathrm{Tr}(\psi^\dagger(x,t)\rho\,\psi(y,t'))\,.
\label{eq:SPC:1-h}
\end{eqnarray}
\end{subequations}
They describe the single-electron coherence
considered in many-body approaches to the problem of electronic coherence in metals\cite{Golubev:1999-3,vonDelft:2008-1}. They are also directly related to
detection signals since the tunneling current from the conductor to a reservoir contains two contributions arising from electron transmitted from
the conductor to the reservoir and vice versa:
\begin{equation}
I_{D}(t)=\int_{0}^t \left(\mathcal{G}_{\rho}^{(e)}(x,\tau;x,\tau')K_{a}(\tau-\tau')-\mathcal{G}_{\rho}^{(h)}(x,\tau;x,\tau')K_{e}(\tau-\tau')\right)\,d\tau d\tau'
\end{equation}
In this expression, $K_{a}(\tau)$ and $K_{e}(\tau)$ characterize the detector and respectively account for available single electron and hole states within the
reservoir and for the eventual energy filtering of the detector. Let us stress that such a detection has been recently implemented experimentally to study
electron relaxation in quantum Hall edge channels\cite{Altimiras:2010-1}.

\medskip

The first important difference with usual quantum optics is the fact that ground state has a non vanishing single particle coherence
whereas Glauber correlators vanish in the photon vacuum. On the contrary, at zero temperature, the single electron coherence within a single chiral 
channel is given by:
\begin{equation}
\label{eq:SPC:2}
\mathcal{G}^{(e)}_{\mu}(x,y) = \frac{i}{2\pi}\,\frac{e^{ik_F(\mu)(x-y)}}{y-x+i0^+}
\end{equation}
where $k_{F}(\mu)$ denotes the Fermi momentum associated with the chemical potential $\mu$ of the edge channel under consideration.
At temperature $T>0~\mathrm{K}$, these correlators decay over the thermal length scale $l(T)=\hbar v_{F} /k_BT$ where $v_{F}$ denotes the
Fermi velocity.

It is then natural to decompose the single electron coherence in a Fermi sea contribution due to the chemical potential $\mu$ of
the conductor and a contribution due to electrons above this ground state:
\begin{equation}
\label{eq:SPC:3}
\mathcal{G}^{(e)}(x,y)=\mathcal{G}^{(e)}_{\mu}(x,y)+\Delta\mathcal{G}^{(e)}(x,y)
\end{equation}
The case of an ideal single electron excitation helps clarifying the physical meaning of
$\Delta\mathcal{G}^{(e)}$. Such a state is obtained from a Fermi sea by adding one 
extra-particle in a normalized wave packet $\varphi_{0}$:
\begin{equation}
\label{eq:SPC:4}
\psi^\dagger[\varphi_{0}]|F\rangle = \int_{-\infty}^{+\infty} \varphi_{0}(x)\,\psi^\dagger (x)\,|F\rangle
\end{equation}
where $|F\rangle $ denotes the Fermi sea at a fixed chemical potential. It is assumed that in
momentum space, $\varphi_0$ only has components on single particle states above the Fermi level. Then, Wick's theorem
leads to
\begin{equation}
\label{eq:SPC:5}
\Delta\mathcal{G}^{(e)}_{\psi^\dagger[\varphi_{0}]|F\rangle}(x,y)=\varphi_{0}(x)\,\varphi_{0}(y)^*\,.
\end{equation}
In the same way, the single electron coherence of 
the state obtained by adding a single hole excitation to the Fermi sea
$\psi[\varphi_{h}]|F\rangle$ ($\varphi_{h}(k)=0$ below the Fermi level) is given by
$\Delta\mathcal{G}^{(e)}_{\psi[\varphi_{h}]\,|F\rangle}(x,y)=
-\varphi_{h}(x)\,\varphi_{h}^*(y)$. Note that the $-$ sign reflects the fact that a 
hole is an absence of an electron from the Fermi sea. These simple examples show that $\Delta\mathcal{G}^{(e)}$ contains all the information relative
to the wavefunctions of excitations emitted by a single electron source.

\section{Quantum tomography}
\label{sec:HBT}

\subsection{Formulating the problem}
\label{sec:HBT:intro}

An important issue of quantum optics is the characterization of the quantum state emitted by a source. Since photons are bosons, 
in the case of a monochromatic source,  the problem is to reconstruct the quantum state of a single mode of the electromagnetic 
field. Spectacular results in quantum state preparation and measurement have recently been achieved in the microwave domain
using cavity QED with Rydberg atoms\cite{Deleglise:2008-1} and superconducting nanocircuits (circuit QED)\cite{Hofheinz:2009-1}. 

On the contrary, in electron quantum optics, the Pauli exclusion principle prevents the emission of more than one electron in a
single particle state. In this context, the question is to characterize the single particle 
excitations emitted by the source: this problem is called single electron tomography and consists in measuring the single
electron coherence $\Delta\mathcal{G}^{(e)}$. This goes beyond the determination of the average current which only probes the
diagonal of $\Delta\mathcal{G}^{(e)}$ in the time domain. This also goes beyond measuring the electron distribution number
since this quantity characterizes single electron coherence only in a stationary situation but not in the case of a pulsed single electron source.

\subsection{The Hanbury Brown \& Twiss effect}
\label{sec:HBT:HBT}

In quantum optics, quantum tomography can indeed be achieved using two particle quantum interferences\cite{Smithey:1993-1,Lvovsky:2009-1}. 
The idea is to rely on the Hanbury Brown Twiss (HBT) effect which consists in intensity correlations and anti-correlations in a coherent scattering 
experiment involving two incoming beams of identical quantum particles. Whereas natural in classical electromagnetism, the effect
was somehow disputed in the optical domain. The controversy stopped with the demonstration of the effect in the laboratory\cite{Hanbury:1956-2,Twiss:1957-1} after
it was used to measure the diameter of stars within the principal sequence\cite{Hanbury:1956-1}. Later,
Fano provided the proper quantum interpretation in term of two photon interferences\cite{Fano:1961-1}. 
Since its discovery, the HBT effect has also been observed for electrons in a metal issued by two different 
reservoirs at equilibrium\cite{Henny:1999-1,Oliver:1999-1,Liu:1998-1}.
It has also been used in nuclear physics where it provides information on the space time geometry of particle pair production regions in 
heavy ion collisions\cite{Baym:1998-1}. More recently, a beautiful experiment conducted with metastable Helium atoms 
demonstrated the effect for fermionic (${}^3\mathrm{He}$) and bosonic 
(${}^4\mathrm{He}$) atoms which exhibited the bunching and antibunching behavior expected from their quantum statistics\cite{Jeltes:2007-1}.

\medskip

\begin{figure}
\begin{center}
\includegraphics[width=10cm]{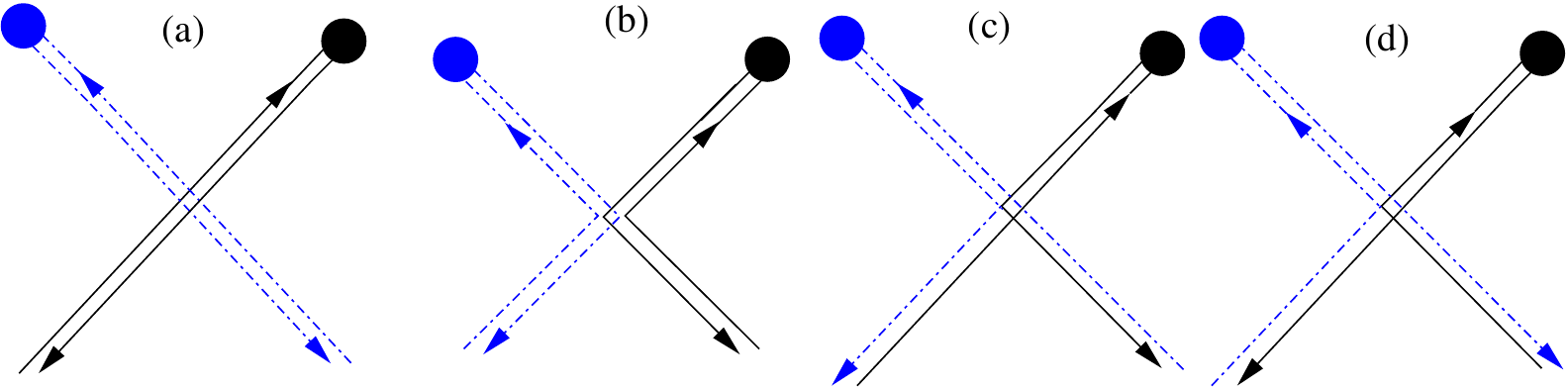}
\end{center}
\caption{\label{fig:HBT} Contributions to the probability of coincidence detection in quantum mechanics: the dashed (resp. full) lines 
corresponds to a single particle ending in the left (resp. right) detector represented here as filled blue (resp. black) dots. 
Each ascending arrows corresponds to a quantum amplitude associated with
a single particle trajectory whereas a descending arrow
corresponds to the complex conjugate of such an amplitude. Diagrams (a) and (b) thus represent classical two particle trajectories whereas (c) and (d) represent
contributions of interferences between direct and exchange paths for two particle trajectories which are responsible for the
HBT effect.}
\end{figure}

To understand the effect, imagine that we observe a coincidence between 
two detectors $A$ and $B$ observing an ensemble of sources, each of them emitting a single particle for simplicity. 
Figure \ref{fig:HBT} depicts the various contributions to such a coincidence event. Two of them can be interpreted in terms
of classical trajectories where a particle emitted by a given source is detected by a given detector. But for indistinguishable quantum particles,
the coincidence probability involves interferences between two particle paths which differ by exchange. These are thus sensitive to the particle's quantum statistics.

Now if we imagine having single particle sources, the HBT effect will occur if and only if both sources emit particles in the same quantum state
and in a synchronized way otherwise they could be distinguished\cite{Degio25}. This has been observed for two photons 
emitted by a non linear cristal by Hong, Ou and Mandel (HOM)\cite{Hong:1987-1} and 20 years later
for photons emitted by two independent sources\cite{Beugnon:2006-1}.

\medskip

This strongly suggests that a controlled source could be used to characterize the single particle content of an unknown source using the HBT
interferometry setup depicted on Figure \ref{fig:tomography}. This idea has been exploited to perform the quantum tomography of a single mode of the electromagnetic
field using a laser as the known source, called a local oscillator\cite{Smithey:1993-1}. 
But in the context of electron quantum optics, Fermi statistics forbids the existence of classical wave and thus
of a laser.  For single electron tomography, we need a controlled and characterized 
source able to sweep the space of single particle states. An ohmic contact provides this appropriate local oscillator since its chemical potential can 
be varied to scan the relevant energy range of single electron and hole excitations emitted by the source to be characterized.

\subsection{Single electron quantum tomography}
\label{sec:HBT:tomography}

\begin{figure}
\begin{center}
\includegraphics[width=10cm]{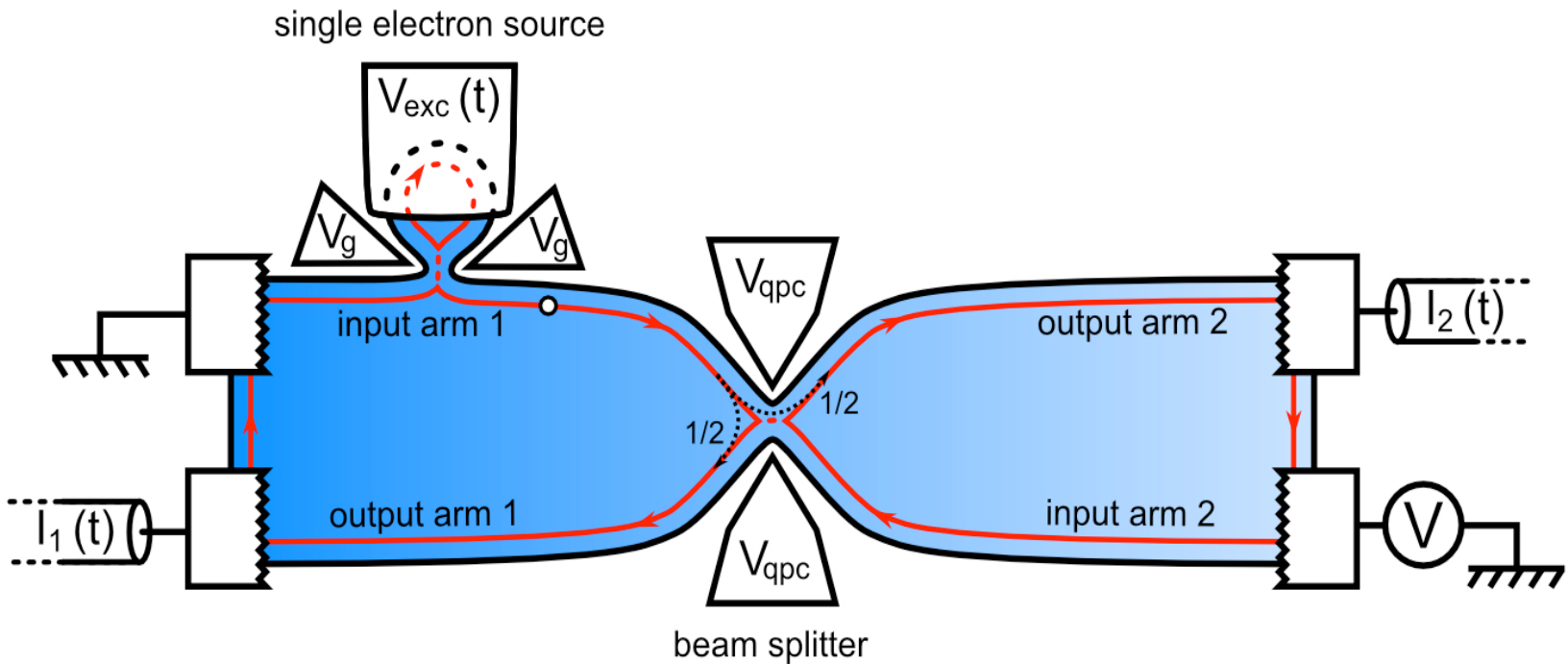}%
\end{center}
\caption{\label{fig:tomography} Hanbury Brown \& Twiss setup for single electron tomography:
This setup is designed to characterize the single electron coherence of an on demand single electron source present on the incoming branch 1
close to the QPC (here $\mu_{1}=0$) and driven by the voltage $V_{\mathrm{exc}}(t)$. A reservoir with a time dependent chemical 
potential $\mu_{2}(t)=-eV(t)=\mu_{2}-eV_{\mathrm{ac}}(t)$ is connected to the incoming branch 2. One measures the low frequency correlation 
$S_{12}^{\mathrm{exp}}$ of the outcoming current $I_{1}$ and $I_{2}$.}
\end{figure}

The Hanbury~Brown \& Twiss (HBT) setup depicted on figure \ref{fig:tomography} is designed to perform
the analogous of homodyne tomography in quantum optics. A quantum point contact (QPC) playing the role of a beam splitter for electrons.
The source to be characterized is placed on the incoming channel 1 whereas an Ohmic contact with controlled
time dependent chemical potential $\mu_{2}(t)=\mu_{2}-eV_{\mathrm{ac}}(t)$ is placed on incoming channel 2. We measure current 
correlations between the two outcoming channels. Time dependent
correlations are defined as:
\begin{equation}
S_{\alpha\beta}^{\mathrm{out}}(t,t')=\langle i^{\mathrm{out}}_{\alpha}(t)\,i^{\mathrm{out}}_{\beta}(t')\rangle -
\langle i^{\mathrm{out}}_{\alpha}(t)\rangle\langle i^{\mathrm{out}}_\beta(t')\rangle\,.
\end{equation}
The outcoming current correlations can be expressed in terms 
of the incoming ones $S_{\alpha\beta}^{\mathrm{in}}(t,t')$ and of a contribution $\mathcal{Q}(t,t')$ coming from two particle interferences. 
This HBT contribution $\mathcal{Q}(t,t')$ involves the incoming single particule coherences right
upstream the QPC:
\begin{equation}
\label{eq:Q}
\mathcal{Q}(t,t')=(ev_F)^2\left(
\mathcal{G}_{1}^{(e)}\mathcal{G}_{2}^{(h)}+
\mathcal{G}_{2}^{(e)}\mathcal{G}_{1}^{(h)}\right)(t',t)\,
\end{equation}
and the final expressions for outcoming current correlations are
\begin{eqnarray}
\label{eq:QPC:S11}
S_{11}^{\mathrm{out}}(t,t') & = & \mathcal{R}^2S_{11}^{\mathrm{in}}(t,t')+\mathcal{T}^2S_{22}^{\mathrm{in}}(t,t') +\mathcal{RT}\,\mathcal{Q}(t,t')\\
\label{eq:QPC:S22}
S_{22}^{\mathrm{out}}(t,t') & = & \mathcal{T}^2S_{11}^{\mathrm{in}}(t,t')+\mathcal{R}^2S_{22}^{\mathrm{in}}(t,t') +\mathcal{RT}\,\mathcal{Q}(t,t')\\
\label{eq:QPC:S12}
S_{12}^{\mathrm{out}}(t,t')&  = & S_{21}^{\mathrm{out}}(t,t') = \mathcal{RT}\,(S_{11}^{\mathrm{in}}(t,t')+S_{22}^{\mathrm{in}}(t,t')-\mathcal{Q}(t,t'))\,
\end{eqnarray} 
where $\mathcal{R}$ and $\mathcal{T}$ are the reflexion and transmission probabilities of the QPC.
Equation \eqref{eq:Q} shows that the HBT contribution is of truly quantum origin: it vanishes when the time separation
is larger than the quantum coherence time within the two incoming channels. Outcoming correlations are then caused by
classical partitioning of current fluctuations at the QPC. To perform single electron quantum tomography
we reconstruct $\Delta\mathcal{G}_{1}^{(e)}$ for the unknown source using \eqref{eq:Q} by studying the variation of the HBT contribution when parameters
of the controlled source on channel 2 are varied. 

The quantum tomography protocols relies on the measurement of the time averaged low frequency component of current correlations:
\begin{equation}
S_{\alpha\beta}^{\mathrm{exp}}=\int \overline{S_{\alpha\beta}^{\mathrm{out}}(\bar{t}+\tau/2,\bar{t}-\tau/2)}^{\bar{t}}\,d\tau
\end{equation}
But then, a stationary source on channel 2 can only give access to the stationary part of $\Delta\mathcal{G}^{(e)}_{1}$ since one measures 
time averaged quantities. Applying a periodic voltage to the ohmic contact at
frequencies corresponding to the harmonics of the driving frequency of the source precisely performs the
homodyning needed to capture the $\bar{t}=(t+t')/2$ dependence of $\Delta\mathcal{G}^{(e)}_{1}$.
Details of the tomography protocol are given in our recent work\cite{Degio:2010-4}. Let us just recall 
that the output of our tomography protocol is the single particle coherence $\Delta\mathcal{G}_{1}^{(e)}$
in the frequency domain since current correlations themselves are measured in the frequency domain. 

\medskip

To support the feasability of this tomography protocol, we have made explicit predictions for the single electron coherence and
for the correlation signals expected for an on demand single electron source\cite{Degio:2010-4}. These predictions, obtained within the framework
of Floquet scattering theory, shed light on the various regimes of the source and on its potential performances. They also give estimates
of noise sensitivity required to perform the experiment: $10^{-30}\ \mathrm{A}^2/\mathrm{Hz}$. Such a resolution is within reach
of ultrahigh sensitivity noise measurement techniques\cite{Glattli:2009-1,Parmentier:2010-1}. 

To conclude this section, let us stress that other experiments based on the HBT interferometer have been proposed such as noise measurements
in an HOM setup\cite{Olkhovskaya:2008-1} and spectroscopy of electron flows\cite{Moskalets:2010-1}. Finally, a recent proposal
aims at producing orbitally entangled pairs of electrons by combining an HBT interferometer with 
two Mach-Zehnder interferometers\cite{Splettstoesser:2009-1}.  We thus hope that experimental groups will soon be able to
demonstrate experiments combining an HBT interferometer with single electron sources.

\section{Electron decoherence and relaxation}
\label{sec:decoherence}

Mach-Zehnder interferometry experiments have demonstrated single electron coherence in quantum Hall edge channels but have also drew attention on 
electron decoherence and its mechanisms. In particular, at filling fraction $\nu=2$, experimental studies have shown that inter-channel Coulomb interactions are
the main source of decoherence for electrons propagating within one edge channel\cite{Roulleau:2007-2,Roulleau:2008-1,Neder:2007-4}. 
Recent experimental studies of electron relaxation\cite{Altimiras:2010-1,LeSueur:2010-1,Altimiras:2010-2} have also shown that 
in this system, single electron excitations are ill defined and this suggests that screened inter-channel Coulomb interactions cannot be considered as weak.

The aforementioned experiments are quantum transport
experiments involving non equilibrium distribution functions. But with the demonstration of a single electron source, the problem of 
quasiparticle relaxation (and decoherence) originally considered by 
Landau in Fermi liquid theory\cite{Nozieres-Pines} could be tested experimentally within quantum Hall edge channels in the near future.
This single quasiparticle relaxation problem is simpler than the relaxation of a general stationary non equilibrium electronic 
state but nevertheless is conceptually important since it probes the nature of quasiparticle in quantum Hall edge channels and
questions the validity of the quantum optics paradigm in these systems.

To address this problem, a unified approach to high frequency transport and to decoherence and relaxation of 
single electron excitations has been developed\cite{Degio:2009-1,Degio:2010-1}. In this approach, we consider their propagation across an
interaction region where possibly strong Coulomb interactions are present or which is capacitively coupled to an
external conductor or circuit behaving as an harmonic environment as shown on figure \ref{fig:scattering}. 

\subsection{Interactions and high frequency transport}
\label{sec:decoherence:plasmon-scattering}

\begin{figure}
\begin{center}
\includegraphics[width=10cm]{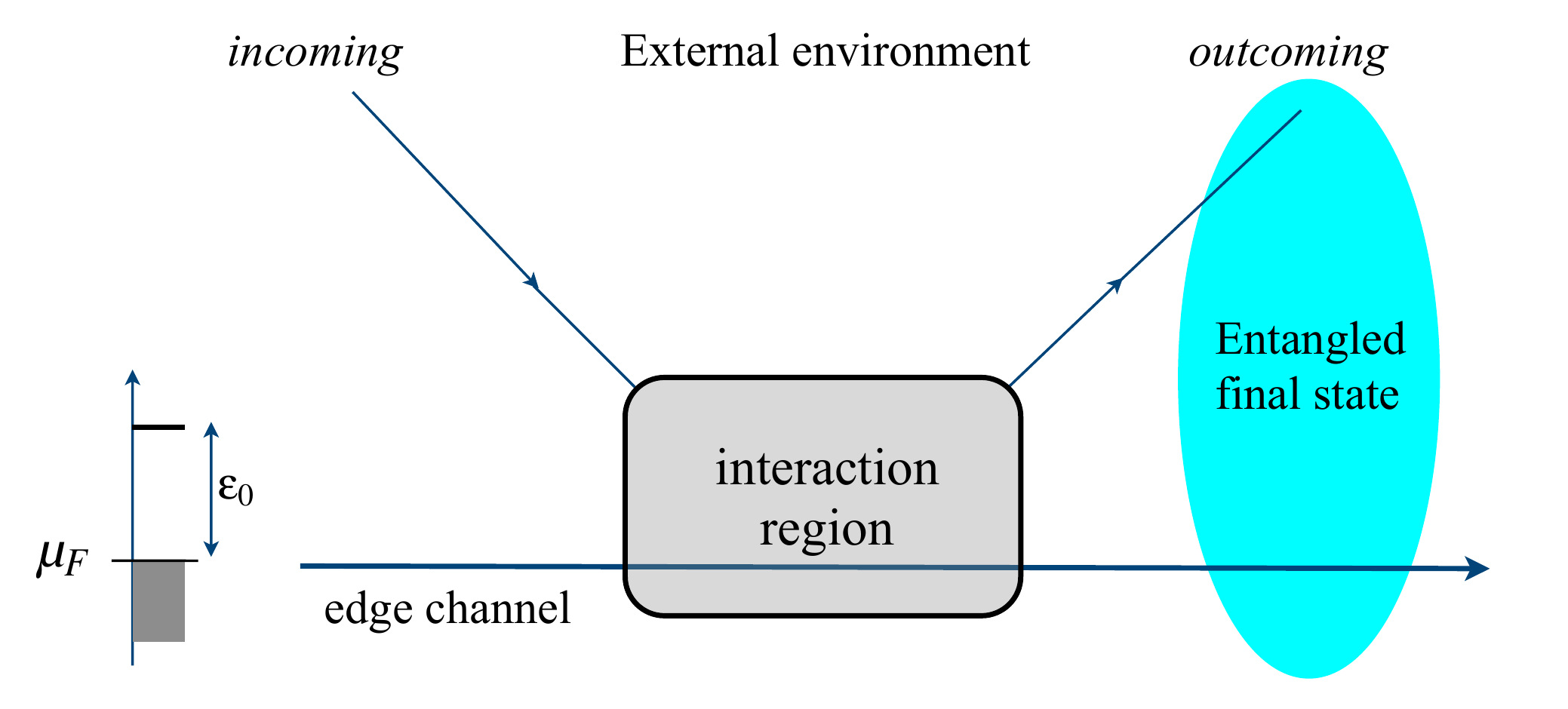}%
\end{center}
\caption{\label{fig:scattering} Scattering approach to decoherence and relaxation in quantum Hall edge channels: an incoming state is injected from
the left into an interaction region in which edge magnetoplasmons and environmental modes are elastically scattered.
This leads to an entangled state between the edge channel and environmental degrees of freedom. We then evaluate 
the outcoming single electron coherence:  $\Delta\mathcal{G}_{\mathrm{out}}^{(e)}(x,y)$.}
\end{figure}

Bosonization essentially describes chiral 1D systems in terms of free bosonic modes, called edge
magnetoplasmons, that describe electron/hole excitations above a Fermi sea vacuum.
Within the bosonization formalism, 
electron/electron interactions as well as capacitive coupling to a linear environment lead to a quadratic
Hamiltonian in the bosonic field describing the bosonized edge channel and the environmental modes.
The equations of motion can thus be solved exactly and, in the geometry of figure \ref{fig:scattering}, it leads to 
a scattering matrix describing the elastic scattering between edge magnetoplasmons
and environmental modes. All the dynamics is encoded into the frequency dependence of the transmission and reflexion coefficients.  

\medskip

Although the plasmon scattering matrix seems to be a rather abstract object, it turns out to be
directly related to finite frequency admittances which are measurable quantities.
Such an idea was pioneered within the context of non chiral 1D Luttinger liquid
by Safi and Schulz\cite{Safi:1995-1,Safi:1995-2} in the mid-90s. 
Let us consider a chiral system involving several edge channels electrostatically interacting within 
an interaction region of length $l$. Each of these channels $\alpha$ is attached to an Ohmic contact driven by a time 
dependent voltage $V_{\alpha}(t)$ generating
a coherent magnetoplasmon state. Within the bosonization formalism, the interaction region acts as frequency 
dependent beam splitter for edge magnetoplasmons. 
The incoming coherent states are then scattered into outcoming coherent states whose amplitudes are given in terms
of the incoming ones by the plasmon scattering matrix, exactly like for classical waves scattered by a beam splitter in classical electromagnetism. 
Using the expression of edge currents in terms of the magnetoplasmon modes,
one immediately obtains a linear relation between the average time dependent currents $I_{\alpha}$ entering the interacting region
of each edge channel in terms of the voltage applied to the Ohmic reservoirs:
\begin{equation}
I_{\alpha}(\omega)=\frac{e^2}{h}\sum_{\beta}g_{\alpha\beta}(\omega)\,V_{\beta}(\omega)\,.
\end{equation}
The important point is that the finite frequency dimensionless admittance matrix $g_{\alpha\beta}(\omega)$ is given in terms of the plasmon scattering matrix
$S_{\alpha\beta}(\omega)$ relating incoming to outcoming plasmonic modes:
\begin{equation}
g_{\alpha\beta}(\omega)=\delta_{\alpha\beta}-S_{\alpha\beta}(\omega)\,.
\end{equation}
For example, in the case of free propagation along a single edge channel with time of flight $\tau$, $S(\omega)=e^{i\omega\tau}$
and therefore $g(\omega)=1-e^{i\omega\tau}$. 
For an edge channel $\alpha$, the relation $g_{\alpha\alpha}(\omega)=1-S_{\alpha\alpha}(\omega)$ is indeed true even 
in the presence of an external electromagnetic environment, provided $S_{\alpha\alpha}(\omega)$ denotes the edge magnetoplasmon transmission amplitude
taking into account scattering into environmental degrees of freedom. Of course, for a given circuit design, this quantity is model
dependent since it requires an explicit description of Coulomb interactions and screening effects. Nevertheless, in the limit where the frequency is
low compared to the inverse of the longest time of flight across the system, a discrete element approach developed by 
B\"uttiker and his collaborators\cite{Buttiker:1993-1,Pretre:1996-1} can be used. In this regime, the low frequency expansion of this finite frequency 
admittance $g(\omega)$ is generically of the form
\begin{equation}
\label{eq:lowfreq}
g(\omega)=-iR_{K}C_{\mu}\omega + \frac{R_q}{R_K}(R_{K}C_{\mu}\omega)^2+\mathrm{higher\ orders}\,.
\end{equation}
where $C_{\mu}$ denotes the electrochemical capacitance and $R_{q}$ the relaxation resistance
of the conductor formed by the interaction region coupled to its environment. As shown by B\"uttiker, Prêtre and 
Thomas\cite{Buttiker:1993-1,Pretre:1996-1},
the relaxation resistance $R_{q}$ is the serial addition of the edge channel relaxation resistance $R_{K}/2$
and of the environmental resistance $R$. The latter is related to the density of states available for dissipation in the
electromagnetic environment of the edge channel. 
The $R_{K}/2$ value is robust to interactions 
within the quantum capacitor's plate\cite{Nigg:2006-1,Mora:2010-1}. It indeed
reflects the presence of a single mode coherent conductor connected to free electron reservoirs since $R_{q}$ is renormalized 
in the fractional quantum Hall regime\cite{Hamamoto:2010-1} and
is equal to the classical result in the presence of strong decoherence effects within the capacitor\cite{Nigg:2008-1}.
The electrochemical capacitance is also the series addition
of its quantum capacitance associated with the dot density of states and of the geometric 
capacitance between the edge channel and its environment. In the case of free propagation, 
expanding $g(\omega)=1-e^{i\omega\tau}$ in powers of $\omega$ and comparing to \eqref{eq:lowfreq} shows that
$R_{K}C_{\mu}$ is nothing but the electron time of flight $\tau$ as noticed by Imry and his collaborators\cite{Ringel:2008-1}. 

\medskip

The B\"uttiker, Prêtre and Thomas predictions have been experimentally tested for the quantum RC circuit 
\cite{Gabelli:2006-1} and also for the quantum RL circuit\cite{Gabelli:2007-1}. As will become clear in the forthcoming section, the availability of 
admittance measurements in the GHz frequency range opens exciting perspectives for quantitative tests of the plasmon scattering approach to
single electron relaxation within quantum Hall edge channels.

\subsection{Quasiparticle relaxation in quantum Hall edge channels}
\label{sec:decoherence:relaxation}

The problem considered by Landau to lay the foundations of Fermi liquid theory\cite{Nozieres-Pines} 
is to determine how a single electron excitation of given energy introduced in an interacting electronic system
relaxes. In the geometry of figure \ref{fig:scattering}, one would consider an incoming single electron excitation 
of momentum $k_{0}>0$ and ask what would be the single electron coherence coming out from the interaction
region.  Because an energy resolved single electron excitation can be described as a superposition of coherent states 
of edge magnetoplasmons, this problem can be solved exactly. The result is a non perturbative solution of the Landau quasiparticle
relaxation problem in quantum Hall edge channels\cite{Degio:2009-1}. 

\medskip

\begin{figure}
\begin{center}
\includegraphics[width=10cm]{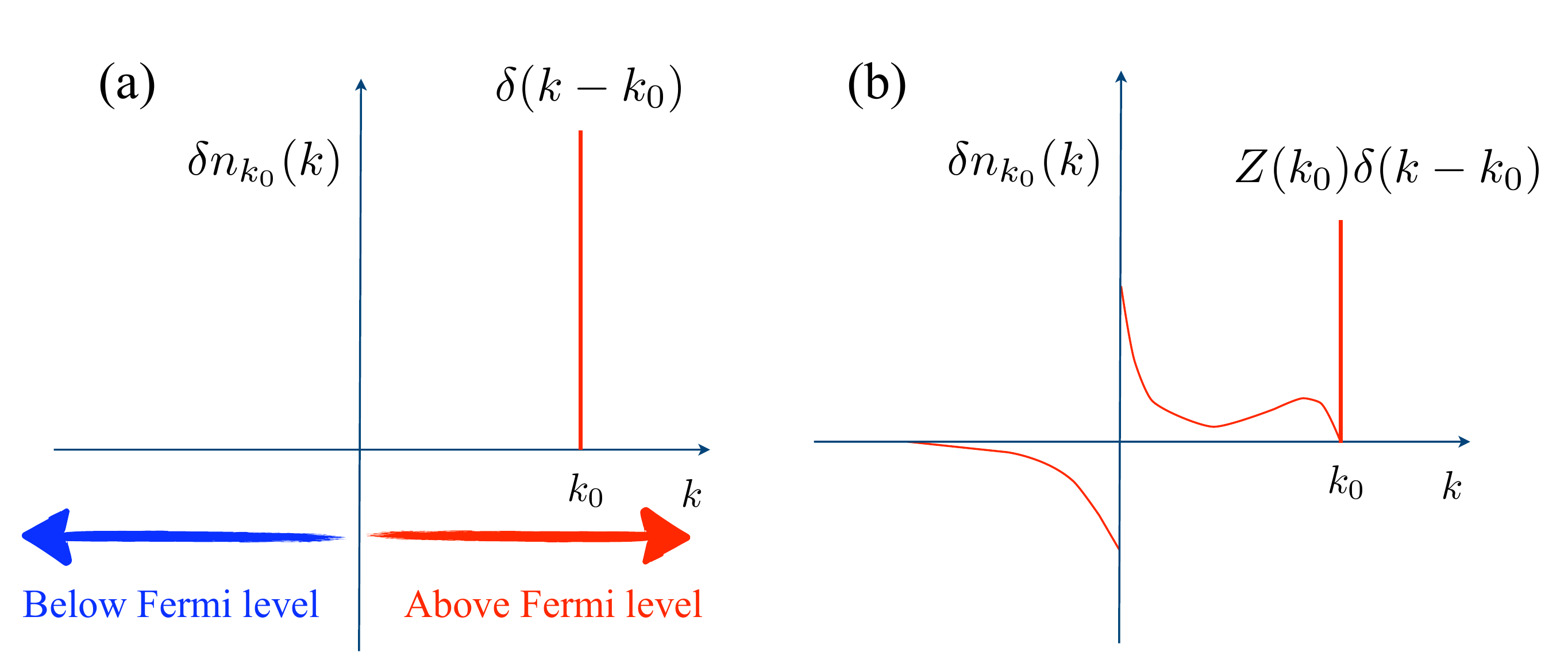}
\end{center}
\caption{\label{fig:relaxation} (a) Incoming single electron coherence for an initial energy resolved single electron excitation above the Fermi level. (b) Outcoming
single electron coherence $\delta n_{k_0}(k)$: at $k=k_0$, the quasiparticle peak comes is the contribution of electrons flying across the interaction region
without loosing energy (elastic scattering probability).}
\end{figure}

At zero temperature, the final result is expected to be of the form depicted on figure \ref{fig:relaxation}:
the incoming electron distribution function $\delta n^{(\mathrm{in})}_{k_{0}}(k)=\delta(k-k_{0})$ comes out of the interacting region
as $\delta n^{(\mathrm{out})}_{k_{0}}(k)=Z(k_{0})\delta(k-k_{0}) + \delta n^r_{k_{0}}(k)$ where $0\leq Z(k_{0})\leq 1$ represent
the probability for the incoming electron to be scattered elastically and
$ \delta n^r_{k_{0}}(k)$ is a regular function, representing the energy relaxation of the incoming electron. Particle conservation implies
the sum rule $\int \delta n^{(\mathrm{out})}_{k_{0}}(k)\,dk=1$. Note that $\delta n^r_{k_{0}}(k)=0$ for $k>k_{0}$ and $k<-k_{0}$ since
we are at zero temperature (no heating) and the available energy from the incoming electron is at most $\varepsilon_{0}=\hbar v_{F}k_{0}$.

\medskip

Our main results\cite{Degio:2009-1} are non perturbative analytical expressions for $Z(k_{0})$
and $\delta n^r_{k_{0}}(k)$ which only depend on the
finite frequency admittance $g(\omega)$. As expected, for  $g(\omega)=1-e^{i\omega\tau}$ (free electron propagation)
$Z(k_{0})=1$ and $\delta n^r_{k_{0}}(k)=0$ meaning that the electron does not relax at all. When interactions are
present, it is interesting to discuss the full non perturbative results in the limit of very low energy to test the validity of the Fermi liquid paradigm 
and, in the limit of increasing energy, to discuss the validity of the quantum optics paradigm for electronic flying qubits\cite{Ionicioiu:2001-1}. 

\begin{figure}
\begin{center}
\includegraphics[width=7.5cm]{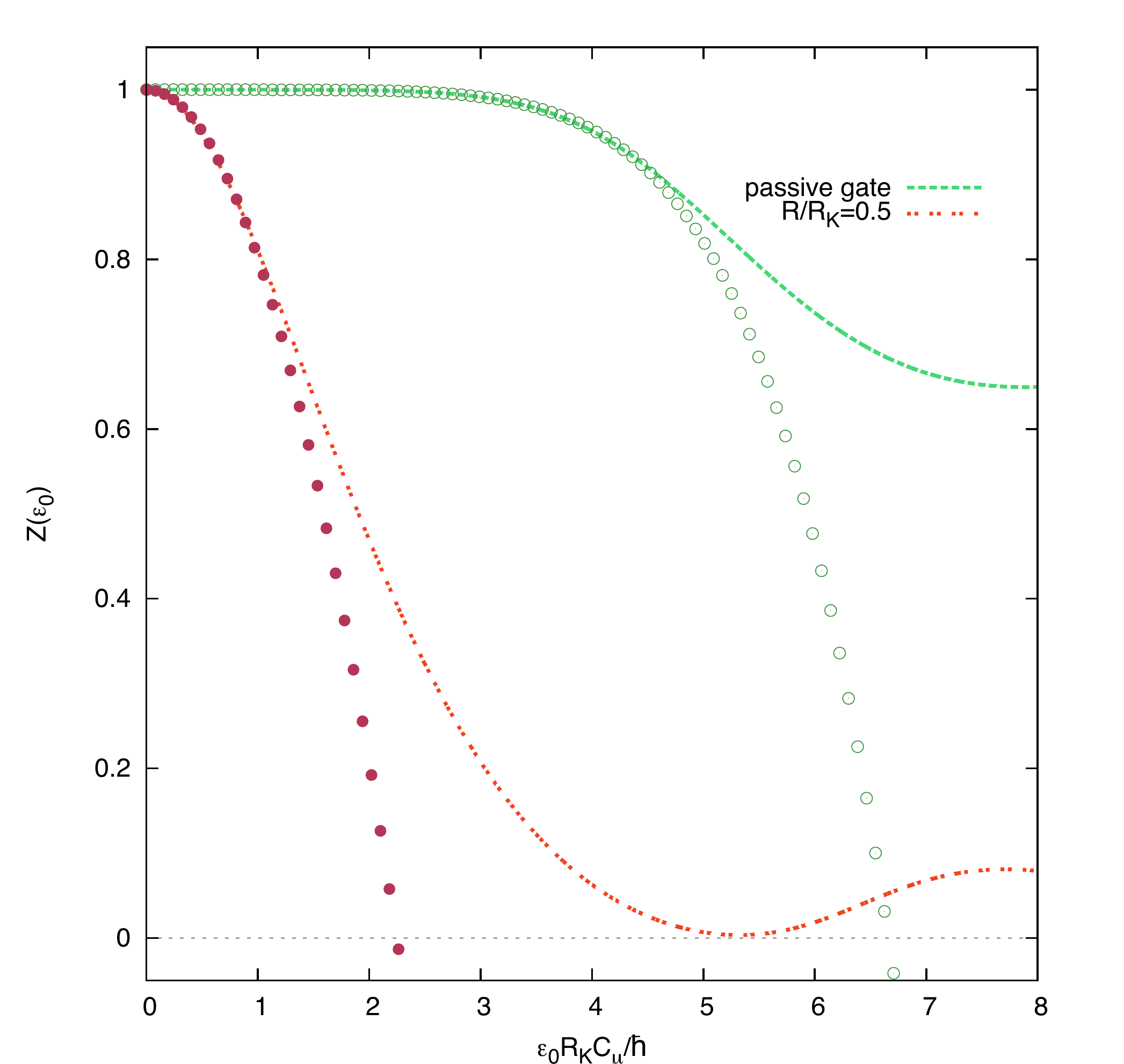}
\end{center}
\caption{\label{fig:low-energy} Elastic scattering probability $Z(\varepsilon_0)$ as a function of 
$\varepsilon_{0}R_{K}C_{\mu}/\hbar$ in the case of the quantum RC circuit for $R=R_K/2$ (dot-dashed line) and
$R=0$ (dashed line) and low energy behaviors given by eqs. \eqref{eq:inelastic:resistive} and \eqref{eq:inelastic:capacitive} (circular dots).}
\end{figure}

At low energy, the Pauli principle blocks inelastic processes and the elastic scattering probability $Z(k_{0})\rightarrow 1$
when $k_{0}$ decays to zero. Using the low frequency expansion of the finite frequency admittance
\eqref{eq:lowfreq}, the low energy behavior of the inelastic scattering 
probability $\sigma_{\mathrm{in}}(\varepsilon_0)=1-Z(\varepsilon_0/\hbar v_F)$ can be obtained as:
\begin{equation}
\label{eq:inelastic:resistive}
\sigma_{\mathrm{in}}(\varepsilon_{0})=\left(\frac{R_q}{R_K}-\frac{1}{2}\right)\,(\varepsilon_{0}R_{K}C_{\mu}/\hbar)^2+\mathrm{higher\ orders}\,.
\end{equation}
The dominant term represents energy losses into the external environment and has a simple universal form. Since the inelastic probability 
scales at least as $\varepsilon_{0}^2$, the quasiparticle is preserved in the limit of very low energies as
expected in a Fermi liquid. The case $R=0$ describes the coupling to a perfectly conducting top gate beneath which electrons within the
chiral edge channel experience long range Coulomb interactions. In this case, at low
frequency, $\sigma_{\mathrm{in}}(\varepsilon_{0})$ starts with a higher power:
\begin{equation}
\label{eq:inelastic:capacitive}
\sigma_{\mathrm{in}}(\varepsilon_{0})=\frac{11}{25920}\,\left(\frac{C_\mu}{C-C_\mu}\right)^2\,(\varepsilon_{0}R_{K}C_{\mu}/\hbar)^6+\mathrm{higher\ orders}\,.
\end{equation}
This behavior is not universal since it still depends on the geometrical capacitance $C$ coupling the edge channel to the top metallic
gate and not only on the electrochemical capacitance $C_{\mu}$ and the relaxation resistance $R_{K}/2$ which describe
the response properties of the quantum RC circuit formed by the interaction region capacitively coupled to a perfectly
conducting top gate. Figure \ref{fig:low-energy} summarizes these results by depicting the exact elastic probability\cite{Degio:2009-1} $Z(\varepsilon_{0})$ 
and their low energy behaviors given by eqs. \eqref{eq:inelastic:resistive} and \eqref{eq:inelastic:capacitive}. Finally, we would like to point out that tailoring properly the 
electrostatic neighborhood of the edge channel so that it doesn't lead to energy leaks leads to decoherence shielding for single electron 
excitations at least at low enough energy. The validity of this idea has been experimentally demonstrated by F.~Pierre and his collaborators\cite{Altimiras:2010-2}.

At higher energies, two scenarii are possible. At low coupling, that is when $g(\omega)$ does not depart too much from a free
electron behavior, the extra electron relaxes but can still be distinguished by its energy from electron/hole
pairs generated within the Fermi sea as illustrated on figure \ref{fig:relaxation}b. 
Then, the Fermi sea can be viewed as an extra dissipation channel. 
In this case, the single electron wavepacket decoheres according to
\begin{equation}
\varphi(x)\,\varphi(y)^*\mapsto \mathcal{D}(x-y)\,\varphi(x)\,\varphi(y)^*
\end{equation}
where $\mathcal{D}(x-y)$ is an effective decoherence coefficient representing the action of an effective environment on the 
quasi particle going through the interaction region. Its expression in terms of the finite frequency admittance
is similar to the one obtained in the dynamical Coulomb blockade theory\cite{Devoret:1990-1,Girvin:1990-1,Ingold:1992-1}:
\begin{equation}
\mathcal{D}(x-y)=\exp{\left(\int_{0}^{+\infty}
\Re{(g(\omega))}(e^{-i\omega(x-y)/v_F}-1)\,\frac{d\omega}{\omega}
\right)}\,.
\end{equation}
But at stronger coupling and high enough energy, this description breaks down. In this case, it is no longer possible
to distinguish between the incoming extra electron and electrons stemming from electron/hole pairs excitations created
within the Fermi sea by Coulomb interactions. A typical situation where this happens is the case of quantum Hall
edges at filling fraction $\nu=2$ which involves two copropagating spin polarized channels coupled 
through Coulomb interactions. 

\subsection{Energy relaxation in $\nu=2$ systems}
\label{sec:decoherence:two-channels}

Recently, F. Pierre and his collaborators have studied energy relaxation in quantum Hall edge channels at $\nu=2$. Using a quantum
dot as a spectrometer, they have measured the electron distribution function in one of the 
two edge channels\cite{Altimiras:2010-1}. Their experimental setup has been designed to study the relaxation of a non equilibrium
distribution function generated by a biased quantum point contact after propagation over several lengths ranging from $l=0.8\ \mu\mathrm{m}$ to
$l=30\ \mu\mathrm{m}$. Experimental data\cite{LeSueur:2010-1,Altimiras:2010-2} show that inter-channel 
interaction is responsible for the energy relaxation in this system and
arises from Coulomb interactions that are screened by metallic gates shaping the 2DEG (see Figure 1 of\cite{LeSueur:2010-1}). 
Therefore, this experiment could be used to test the plasmon scattering approach to interacting quantum Hall edge channels. 

\medskip

In a recent work\cite{Degio:2010-1}, we have studied the
simplest model describing a system of two copropagating quantum Hall edge channels. We have shown, that under physically reasonable assumptions,
the plasmon scattering matrix had a universal form at low energy. This $2\times 2$ unitary matrix $S(\omega,l)$ corresponding to the scattering by an interacting
region of length $l$ is of the form
\begin{equation}
\label{eq:Smatrix}
S(\omega,l)=e^{i\omega l/v_0}
\,e^{-i(\omega l/v)(\cos{(\theta)}\sigma^z+\sin{(\theta)}\sigma^x)}
\end{equation}
where $v_{\pm}^{-1}=v_{0}^{-1}\mp v^{-1}$ are the velocities of the true eigenmodes of the coupled edge channel system which are linear combinations,
characterized by the angle $\theta$, of the edge magnetoplasmon modes of the two channels. Strongly coupled
edge channels correspond to $\theta=\pi/2$, a situation already considered in the recent theoretical study of MZI at $\nu=2$ by 
Levkivsky and Sukhorukov\cite{Levkivskyi:2008-1}. Then, the eigenmodes
of the coupled system correspond to charge and spin density waves, the latter being the slowest of the two. 

\begin{figure}
\begin{center}
\includegraphics[width=12cm]{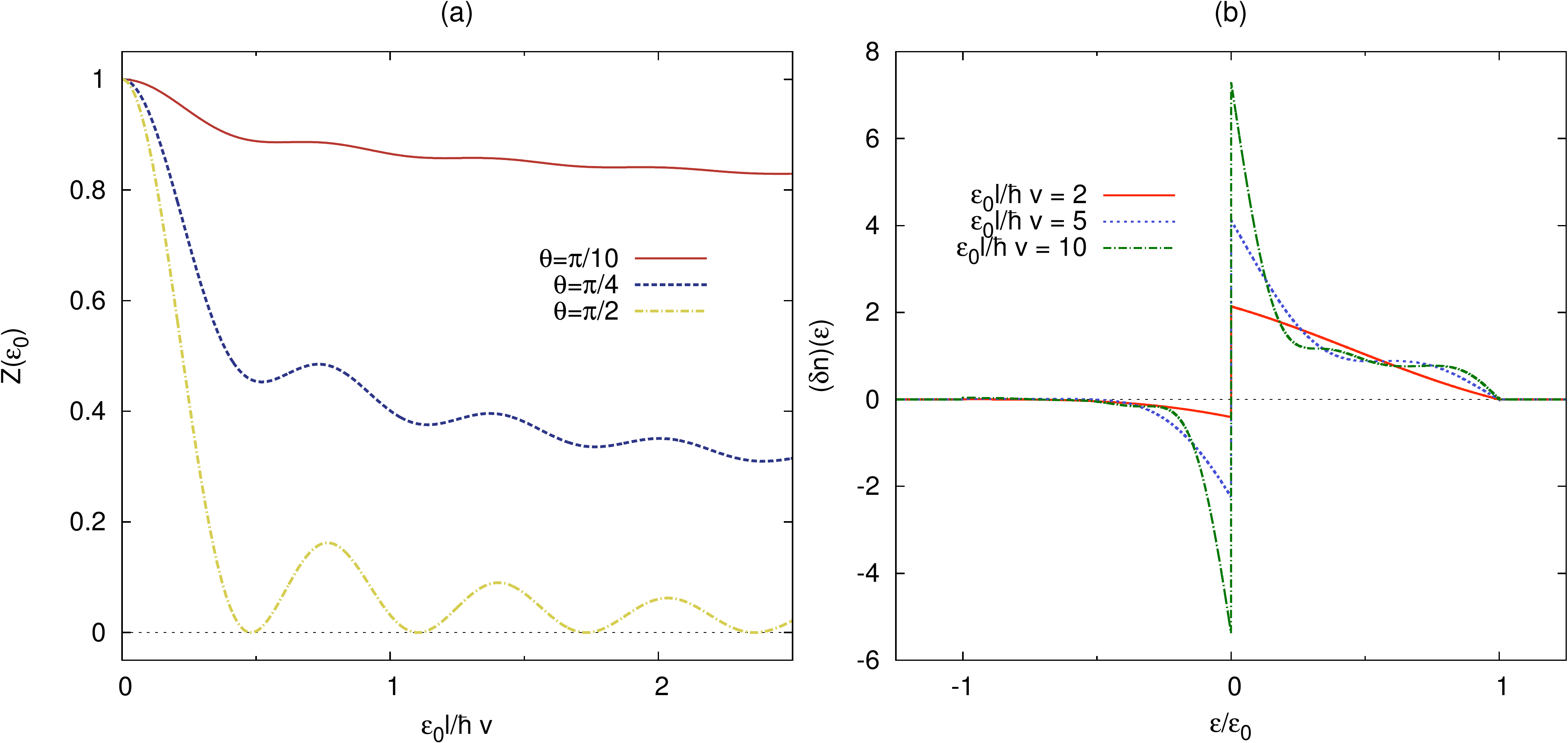}
\end{center}
\caption{\label{fig:relaxation2channels} Relaxation in the two channel system: (a) $Z(\varepsilon_{0})$ as a function of $\varepsilon_{0}l/\hbar v$ 
for weak to strong coupling: $\theta=\pi/10$ (full line), $\theta=\pi/4$ (dashed line), $\theta=\pi/2$ (long dashed line). (b) $\delta n_{\varepsilon_0}^r(\varepsilon)$ as a
 function of $\varepsilon/\varepsilon_{0}$ at strong coupling ($\theta=\pi/2$) for various propagation lengths.}
\end{figure}

Using the scattering matrix \eqref{eq:Smatrix}, the relaxation of single electron excitations can be computed. 
At strong coupling ($\theta=\pi/2$), $Z(\varepsilon_{0})=(J_{0}(\varepsilon_{0}l/\hbar v))^2$
showing that the quasi particle decays after propagation
over $l_{\mathrm{in}}\simeq \hbar v/\varepsilon_{0}$, consistent with the semi-quantitative conclusion from the experimental group\cite{LeSueur:2010-1}. 
Figure \ref{fig:relaxation2channels}a presents the decay of $Z(\varepsilon_{0})$ for various values of $\theta$, showing that $\theta=\pi/2$ leads to the fastest decay
and that, for $\theta\ll \pi/2$ (weak coupling), the quasi particle propagates over a much longer distance.
Looking at the complete electronic relaxation for $\theta=\pi/2$ depicted on figure \ref{fig:relaxation2channels}b shows that, once the quasi particle peak has decayed,
the incident electron has been drawn into the electron/hole pairs it has generated. Thus, although for a fixed $l$, the notion of a quasi particle in this $\nu=2$
edge channel system makes sense at low energy ($\varepsilon_{0}\ll \hbar v/l$), it cannot be viewed as a Fermi liquid in the thermodynamic limit since 
a quasi particle of given energy will disappear after propagation over $l_{\mathrm{in}}(\varepsilon_{0})\simeq \hbar v/\varepsilon_{0}$, meaning that its 
energy width is proportional to its energy. At high energies, relaxation of single electron excitations cannot be described
within a simple dynamical Coulomb blockade picture and electronic quasi particles are ill defined. 

Although computing the exact relaxation of the electron distribution function is a complicated task, it is much simpler
to compute the evolution of heat currents along each edge channel since it only requires two point functions of the current 
injected by the biased QPC. These predictions can then be compared to the experimental data\cite{Degio:2010-1}. We have
shown that using a typical value for the velocity $v\simeq 10^5\ \mathrm{m}\mathrm{s}^{-1}$, the two channel model reproduces
correctly the dependence of the heat current in the QPC bias voltage and length assuming ad hoc 25~\% of the 
energy leaks out. This leaking was interpreted as a manifestation of extra degrees
of freedom such as neutral modes arising from the eventual compressibility of the inner edge channel. 

Note that an alternative approach based on the Boltzmann equation has been developed by Lunde, Nigg and B\"uttiker\cite{Lunde:2010-1} which
also fits the experimental data at the price of assuming also an energy leak towards other degrees of freedom. Finally,
in their recent work\cite{Kovrizhin:2010-2}, Kovrizhin and Chalker have suggested that the effective temperature
deduced from tunneling measurements differs from the effective temperature representing the real heat current due to an interaction induced renormalization
of the tunneling density of state.

At this point, it should be noticed that the total heat current being an integrated quantity, it does not give an energy resolved information. The
electron distribution function does give access to such an information but the issue of the tunneling DOS as well as the difficulty to perform
predictive computations for this quantity implies that it may not be the best quantity to confront theoretical models to experimental data. Within the
plasmon scattering approach, finite frequency current noise directly provides a frequency resolved information on the population of the edge
magnetoplasmon modes. This is why we have suggested to use such measurement as well as finite admittance measurements to probe
edge magnetoplasmon scattering\cite{Degio:2010-1}. 

\medskip

As shown by the flurry of theory papers devoted to the discussion of electron dephasing\cite{Seelig:2001-1,Forster:2005-1} and
fringe visibility\cite{Chung:2005-1,Chalker:2007-1,Neuenhahn:2009-1,Neuenhahn:2008-1,Neder:2007-3,Neder:2007-4,Kovrizhin:2009-1,Kovrizhin:2010-3}, 
the Mach-Zehnder interferometer would deserve a full review by itself. We would just like to point out that although $Z(k_{0})$ 
naturally appears as the contrast reduction of interference fringes for energy resolved single
electrons sent into a Mach-Zenhder interferometer, Levkivskyi and Sukhorukov\cite{Levkivskyi:2008-1} have shown that
understanding MZI experiments involving electron flows from reservoirs also requires, beside decoherence effects, to properly consider
electrochemical potential shifts within the interferometer, a point also stressed by Roche {\it et al}\cite{Roulleau:2008-1}. 
Moreover new phenomena have been predicted:
when a non equilibrium Fermi double step electron distribution function is injected into an incoming channel of the interferometer, non 
Gaussian fluctuations of the current can induce a quantum phase transition on the lobe structure of the visibility\cite{Levkivkyi:2009-1}.

\section{Conclusion and perspectives}
\label{sec:conclusions}

Recent progresses in nanofabrication and radiofrequency technology and
in ultrahigh sensitivity noise measurements have opened the way to new experiments aiming at the controlled manipulation of 
elementary excitations in ballistic quantum conductors similar to what is done with photons in quantum optics.
These experiments have already stimulated an important stream of 
theoretical research on the formulation of  ``electron quantum optics'' taking into account interactions and discussing the 
dynamics of single to few electron excitations in the presence of interactions. 

\medskip

The main message we would like to pass to the reader is that the emerging field of electron quantum optics 
really provides a radically new point of view on quantum transport and more generally on mesoscopic physics. 

\medskip

First of all, electron quantum optics really concerns the exploration of time scales comparable to
the characteristic times of electron dynamics within quantum conductors, {\it i.e.} the time of flight and the decoherence time which
are of the order of tens to hundreds of picoseconds. Next, instead of dealing with non-equilibrium many body states involving a large
number of electrons, electron quantum optics deals with one to few elementary excitations, a situation conceptually much simpler than in a
far from equilibrium quantum transport experiment. Interestingly the recently developped non equilibrium 
bosonization\cite{Gutman:2010-1,Gutman:2010-2} should
provide the technical basis for understanding the role of non Gaussian fluctuations in single electron devices and also for understanding
the close relation between electron quantum optics and full counting statistics\cite{Snyman:2008-1}.

\medskip

Moreover, on the experimental side, electron quantum optics clearly requires a new way of doing mesoscopic physics by bringing
new tools to prepare, control and record quantum states. On the conceptual side, it brings in concepts from quantum information
theory such as entanglement, fidelities and entanglement entropies. 

\medskip

Finally, although analogies with quantum optics should be stressed, a rich physics emerges from differences between photons and electrons:
understanding the interplay of Coulomb interactions and of the Fermi statistics in a mesoscopic conductor is a highly non 
trivial problem which has already attracted a lot of attention\cite{Degio:2010-2,vonDelft:2008-1}.
It touches deep questions such as the nature and dynamics of quasiparticles in mesoscopic conductors.
Electron quantum optics can bring new insights on these important issues since we can now envision {\it gedanken experiment} never 
realized before such as controlled decoherence experiments involving single to few electron excitations
by exploiting controlled couplings to external circuits. Our highest hopes are therefore within the hands of experimentalists. The 
successful demonstration of new electron quantum optics experiments such as single electron tomography 
will really mark the beginning of a new era.

\acknowledgements{We would like to thank C. Altimiras, M. Albert, J.M. Berroir, E. Bocquillon, M.~B\"uttiker, Ch. Flindt,
Ch. Glattli, Th. Jonckheere, H. Le Sueur, I. Levkivskyi, Th. Martin, S. Nigg, F. Pierre, F. Parmentier, B. Pla\c{c}ais, 
F. Portier, J.M. Raimond, J. Rech, P. Roche, P. Roulleau and E. Sukhorukov for useful discussions.}

\end{document}